\documentclass[iop]{emulateapj}
\usepackage{graphicx}
\usepackage{hyperref}
\hypersetup{colorlinks=True,pdfborder={0 0 0},breaklinks,colorlinks,citecolor=blue,linkcolor=magenta}
\usepackage{natbib}
\usepackage{float}

\newcommand{{\HST}}{\emph{HST}}
\newcommand{{\M}}{log($M_{*}$/M$_{\odot}$)}
\newcommand{{\m}}{$m_{0}$}
\newcommand{\kpc}{{~kpc}}
\newcommand{{\n}}{S\'{e}rsic}
\newcommand{\zr}{$1\textless{z}\textless7.2$}
\newcommand{\ms}{mass$-$size}
\newcommand{\msr}{mass$-$size relation}
\newcommand{\s}{~M$_{\odot}$}
\newcommand{\sfgs}{star-forming galaxies}
\newcommand{\ack}{Research support to R.J.A is provided by the Australian Astronomical Observatory. 
K.G. acknowledges the support of the Australian Research Council through Discovery Proposal awards DP1094370, DP130101460, and DP130101667. 
G.G.K. acknowledges the support of the Australian Research Council through the award of a Future Fellowship (FT140100933). 
Australian access to the Magellan Telescopes was supported through the National Collaborative Research Infrastructure Strategy of the Australian Federal Government. 
We thank the Las Companas Observatory for access to facilities for the ZFOURGE survey.}

\shorttitle{The Evolution of Star-Forming Galaxy Sizes}
\shortauthors{R. Allen et al.}

\begin{document}

\title{The Size Evolution of Star-forming Galaxies Since $z\sim7$ Using ZFOURGE}

\author{Rebecca J. Allen\altaffilmark{1,3},
Glenn G. Kacprzak\altaffilmark{1},
Karl Glazebrook\altaffilmark{1},
Ivo Labb\'{e}\altaffilmark{5},
Kim-Vy H. Tran\altaffilmark{2},
Lee R. Spitler\altaffilmark{3,4},
Michael Cowley\altaffilmark{3,4},
Themiya Nanayakkara\altaffilmark{1},
Casey Papovich\altaffilmark{2}, 
Ryan Quadri\altaffilmark{2},
Caroline M. S. Straatman\altaffilmark{6},
Vithal Tilvi\altaffilmark{2},
Pieter van Dokkum\altaffilmark{7}
}
                                                                                
\altaffiltext{1}{Swinburne University of Technology, Victoria 3122, Australia}
\altaffiltext{2}{George P. and Cynthia Woods Mitchell Institute for Fundamental Physics and Astronomy, and Department of Physics and Astronomy, Texas A\&M University, College Station, TX 77843}
\altaffiltext{3}{Australian Astronomical Observatories, PO Box 915, North Ryde, NSW 1670, Australia}
\altaffiltext{4}{Department of Physics \& Astronomy, Macquarie University, Sydney, NSW 2109, Australia}
\altaffiltext{5}{Leiden Observatory, Leiden University, P.O. Box 9513, 2300 RA Leiden, The Netherlands}
\altaffiltext{6}{Max Planck Institute for Astrophysics, Karl-Schwarzschild-Str. 1, Postfach 1317, D-85741 Garching, Germany}
\altaffiltext{7}{Department of Astronomy, Yale University, New Haven, CT 06520, USA}

\begin{abstract}
For the first time, we present the size evolution of a mass-complete ({\M}$\textgreater10$) sample of {\sfgs} over redshifts $z=1-7$, selected from the FourStar Galaxy Evolution Survey (ZFOURGE).
Observed H-band sizes are measured from the Cosmic Assembly Near-Infrared Deep Extragalactic Legacy Survey (CANDELS) Hubble Space Telescope ({\HST})/F160W imaging. 
Distributions of individual galaxy masses and sizes illustrate that a clear {\msr} exists up to $z\sim7$.
At $z\sim7$, we find that the average galaxy size from the {\msr} is more compact at a fixed mass of {\M}$=10.1$, with $r_{1/2,maj}=1.02\pm{0.29}${\kpc}, than at lower redshifts.
This is consistent with our results from stacking the same CANDELS {\HST}/F160W imaging, when we correct for galaxy position angle alignment.
We find that the size evolution of star-forming galaxies is well fit by a power law of the form $r_e = 7.07(1 + z)^{-0.89}$~kpc, which is  consistent with previous works for normal star-formers at $1\textless{z}\textless4$. 
In order to compare our slope with those derived Lyman break galaxy studies, we correct for different IMFs and methodology and find a slope of $-0.97\pm{0.02}$, which is shallower than that reported for the evolution of Lyman break galaxies at $z\textgreater4$ ($r_e\propto(1 +z)^{-1.2\pm0.06}$).
Therefore, we conclude the Lyman break galaxies likely represent a subset of highly {\sfgs} that exhibit rapid size growth at $z\textgreater4$.
\end{abstract}

\section{Introduction}
The {\ms} and luminosity$-$size relations, have been used to show how {\sfgs} have grown in size since $z\geq6$ \citep[e.g.,][]{2010ApJ...709L..21O,2012ApJ...756L..12M,2015ApJ...808....6H,2015ApJS..219...15S}.
Identifying {\sfgs} at these redshifts and quantifying their growth is paramount for constraining their assembly mechanisms and early mass growth \citep[e.g.,][]{1978MNRAS.183..341W,2001MNRAS.324..313S}. 
Studying the sizes of {\sfgs} at $z\sim6$ is also important because they are progenitor candidates of massive compact quiescent galaxies at $z\sim4$ \citep{2015ApJ...808L..29S}.

Lyman break galaxies (LBGs) have been exclusively used to study galaxy size growth above $z\sim3-4$ \citep[e.g.,][]{2004ApJ...616L..79B}.
However, LBGs are selected via filter dropout techniques \citep{2000ApJ...532..170S}, have bright UV magnitudes, have median masses of {\M}$\textless10$ \citep{2012ApJ...756L..12M}, and do not represent a mass-complete sample of {\sfgs}.
The growth of LBGs appears to be rapid with redshift following $r_e\propto(1+z)^{-1.2}${\kpc} \citep{2016ApJ...821...72S}. 
This rate differs from the size evolution found for mass-complete samples of {\sfgs} at $z\textless4$ where $r_e\propto(1+z)^{-0.75}${\kpc} \citep{2014ApJ...788...28V,2015ApJ...808L..29S}.  
Although there is yet to be a mass-complete galaxy survey that overlaps in the same redshift regime as the LBG studies.

The ZFOURGE survey has provided evidence of galaxy diversity at $z=3-4$, with a high fraction of mature dusty star-formers and quenched galaxies in place \citep{2014ApJ...787L..36S,2014ApJ...783L..14S}.
In fact, \citet{2014ApJ...787L..36S} found that for a mass-complete sample of {\sfgs} ({\M}$\textgreater10.6$), the majority are dusty with a median $A_V$ of $1.7\pm{0.3}$~mag.
These galaxies, as well as the unobscured star-formers in the sample, have UV magnitudes that are at least $\sim5$ times fainter than LBGs, as well as median masses that are much higher. 
Therefore, the entire population of high redshift {\sfgs} are likely not LBGs.

To understand the size evolution, and better constrain the formation and assembly, of a general population of {\sfgs}, a mass-complete analysis is necessary.
In this paper, we analyse the size evolution of a mass-complete ({\M}$\textgreater10$) sample of {\sfgs} to $z\sim7$, using the ZFOURGE survey, and compare it to the evolution of LBGs. 

The paper is organized as follows: in Section~\ref{sec:sam} we describe our sample selection and its properties, in Section~\ref{sec:size} we describe our construction of the mass-size relation and image stacks, followed by our results regarding the average sizes and size evolution in Section~\ref{sec:res}.
We discuss the consequences of our findings in Sections~\ref{sec:dis} and \ref{sec:con}.
We assume a $\Lambda$CDM cosmology with $\Omega_{\Lambda}=0.70$, $\Omega_{m}=0.30$, and $H_{0}=70$ km s$^{-1}$.
\section{Galaxy Sample Selection}
\label{sec:sam}
\begin{figure*}[h]
\epsscale{1.15}\plotone{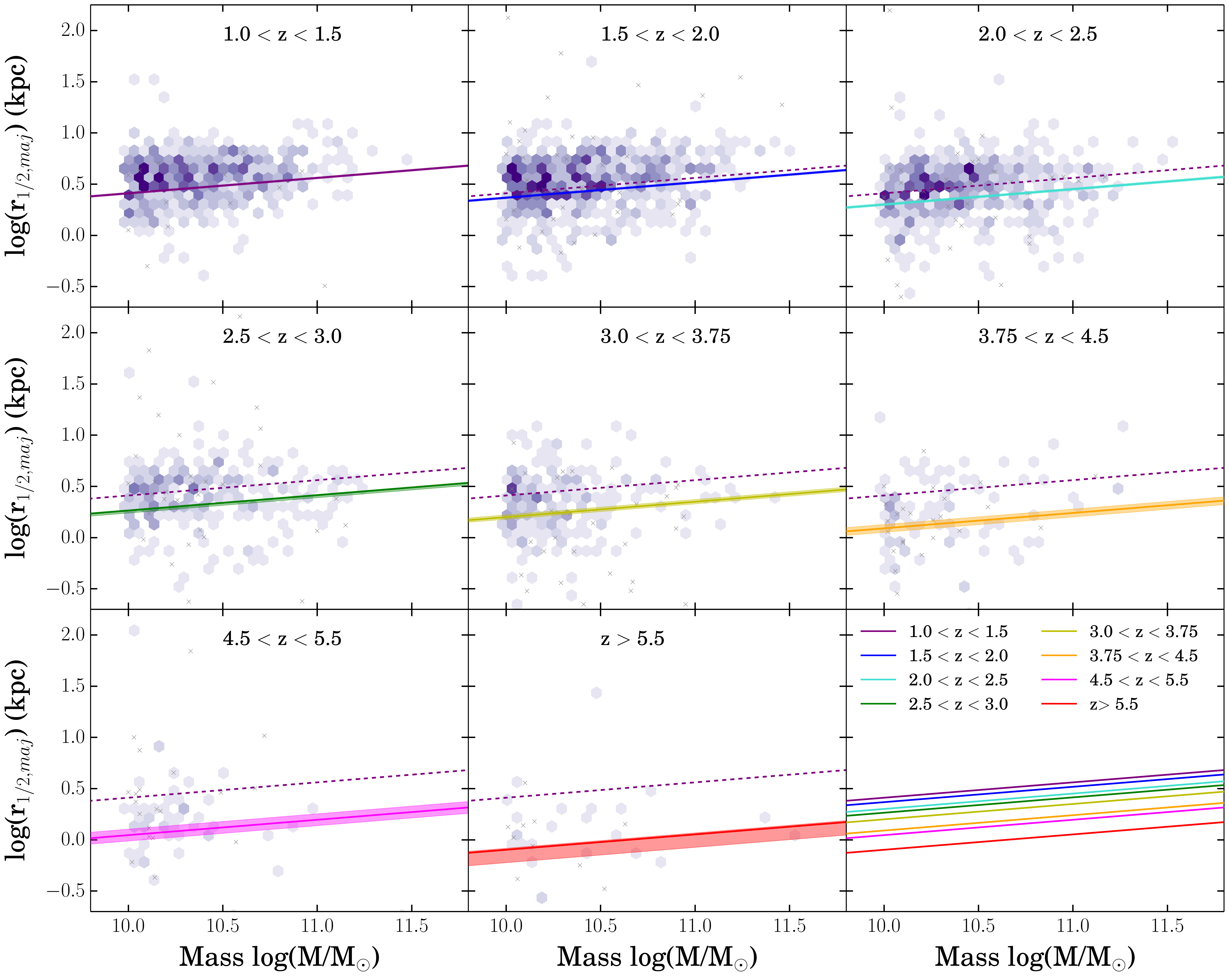}
  \caption{Galaxy stellar {\ms} distributions for {\sfgs} for redshifts {\zr}. In each panel, galaxies with reliable fits from GALFIT are shown as purple points, while galaxies with poor fits are shown as grey crosses, which are not included in the fits. We fit each distribution following $r~(m_{*})/{kpc}=r_0~{\cdot}~{m}_{*}^{0.15}$, which is shown as a contour including the $1\sigma$ errors. The best-fit for $r_{1/2,maj}$ decreases with redshift, which we demonstrate by including the lowest redshift fit in panels with $z\textgreater1.5$ as a dashed purple line. In the last panel, we include the fitted {\msr} from all redshifts using the same colored lines, showing the redshift evolution of galaxy sizes.}
  \label{fig:size}
\end{figure*}
Our sample of galaxies is selected from the ZFOURGE near-IR medium-band survey, which consists of $\sim70,000~K_{s}$-band selected objects \citep{2016ApJ...830...51S} in the COSMOS~\citep{2007ASPC..375..166S}, UDS~\citep{2007MNRAS.379.1599L}, and CDFS~\citep{2002ApJS..139..369G} fields.
The ($5\sigma$) limiting magnitudes of the detection images are 25.5-26.5 AB and are $80\%$ complete at $K_{s}\textless25.3-25.9$.
ZFOURGE takes advantage of the near-IR medium-band filters J1, J2, J3, Hs and Hl on the {\tt{FourStar}} instrument~\citep{2013PASP..125..654P} on the Magellan telescope \citep{2013ApJ...768...56T, 2016ApJ...830...51S}. 
The medium-band filters span a rest-frame wavelength range of $1.05-1.8~\mu$m, which is ideal for following the $4000$~{\AA} break feature of galaxy SEDs from $1.5\textless{z}\textless4$.
The use of medium-bands as well as ancillary photometric data (spanning $0.3-8~\mu$m rest-frame) allow for reliable SED fitting to calculate redshifts, stellar masses, rest-frame colors and other galaxy properties.
Photometric redshifts are calculated with EAZY~\citep{2008ApJ...686.1503B} by fitting the photometric data with stellar population models from~\citet{2003MNRAS.344.1000B}.
Galaxy properties, such as stellar masses, dust extinction, and SFRs, are calculated via SED fitting with FAST~\citep{2009ApJ...700..221K}.
The stellar population models assume a \citet{2003PASP..115..763C} initial mass function and exponentially declining star-formation rate with $\tau=10$~Myr to 10~Gyr. 
For a full description of the survey and data products, see \citet{2016ApJ...830...51S}.

Our sample of {\sfgs} is $80\%$ mass-complete to {\M}$\textgreater10$ (Spitler et al., in prep) at {\zr}.
At $z\textless4$, we use the rest-frame UVJ diagram to separate star-forming and quiescent galaxies \citep{2009ApJ...691.1879W}.
For galaxies above $z=4$, the rest-frame $4000$~{\AA} break feature is redshifted beyond the K-band filter; therefore, we inspect individual SED fits for all galaxies above $z=4$ to confirm that they have reliable photometric redshifts.
While the rest-frame UVJ colors can still be used to identify {\sfgs} at $z\textgreater4$, Spitler et al. (in prep.) find that the fraction of quiescent galaxies drops to essentially zero at masses from {\M}$=10$ at $z=3.5$, therefore we assume that contamination of quiescent galaxies is negligible.
\section{Galaxy Sizes}
\label{sec:size}
We define a galaxy's size as the half-light radius along the semi-major axis, $r_{1/2,maj}$ as measured from GALFIT \citep{2010AJ....139.2097P}.
Galaxy sizes are determined from observed H-band {\HST}/F160W imaging, which is equivalent to rest-frame R to B-band at $z\textless2.5$, and rest-frame U-band above $z=2.5$. 
At $z\textless2.5$, the H-band is equivalent to rest-frame R to B-band and at $z\textgreater2.5$ it is equivalent to rest-frame U-band.  
We do not apply any color-size corrections since color gradients decrease with redshift for {\sfgs} \citep{2014ApJ...788...28V}.
We use both individual galaxy sizes and galaxy image stacks to determine the average sizes of {\sfgs} at fixed mass since $z\sim7$.
\subsection{Individual galaxy sizes from CANDELS and 3D-HST}
Individual galaxy sizes are obtained by cross-matching the ZFOURGE catalogue with the size catalogues of \citet{2014ApJ...788...28V}.
They determine $r_{1/2,maj}$ using GALFIT~\citep{2010AJ....139.2097P} to fit a single component {\n} profile to CANDELS {\HST} imaging \citep{2011ApJS..197...35G,2011ApJS..197...36K} of galaxies in the 3D-HST photometric catalogs~\citep{2014ApJS..214...24S}.
For more information on the fitting procedure and quality of the sizes please see~\citet{2012ApJS..203...24V}.

We use the photometric redshifts, stellar masses, and individual sizes to construct {\ms} distributions in redshift bins to $z\sim7$, Figure~\ref{fig:size}.
Galaxies flagged as having poor fits from GALFIT in the \citet{2014ApJ...788...28V} catalogs are excluded from our analysis; however, these galaxies do not affect our results when included.
The number of poor fits is not mass dependent, but the fraction of good fits decreases from $\sim90\%$ below $z=4$ to $\sim60\%$ at $z\textgreater4$ (see Figure~\ref{fig:size} and Table~\ref{table:fit}).
\begin{table}
\begin{center}
\caption{Best-fit values for $A$, mass-normalised average sizes, and average sizes from image stacks.}
\label{table:fit}
	\begin{tabular}{c c c c c c c}
	\hline\hline
	&&Fit$^*$&&Stack$_{Corr.}^{**}$&\\[0.5ex]
	\hline
	Redshift & $r_0$ &  $r_{1/2,maj}$ & N$_{fit}$ & $r_{1/2,maj}$ & Corr. & N$_{stack}$ \\[0.5ex]
	 &  & (kpc) & & (kpc)&$\%$ \\[0.5ex]
	\hline
	$1.0\textless{z}\textless1.5$ & 0.52 & $3.28\pm{0.07}$ & 542 & $3.43\pm{0.10}$ & 21 & 563 \\[0.5ex]
	$1.5\textless{z}\textless2.0$ & 0.47 & $2.96\pm{0.07}$ & 645 & $3.23\pm{0.12}$ & 22 & 708 \\[0.5ex]
	$2.0\textless{z}\textless2.5$ & 0.41 & $2.55\pm{0.07}$ & 508 & $2.78\pm{0.06}$ & 21 & 584 \\[0.5ex]
	$2.5\textless{z}\textless3.0$ & 0.37 & $2.34\pm{0.13}$ & 254 & $2.56\pm{0.23}$ & 18 & 327 \\[0.5ex]
	$3.0\textless{z}\textless3.75$ & 0.31 & $2.02\pm{0.10}$ & 201 & $1.88\pm{0.10}$ & 22 & 289 \\[0.5ex]
	$3.75\textless{z}\textless4.5$ & 0.20 & $1.57\pm{0.14}$ & 89 & $1.66\pm{0.10}$ & 22 & 146 \\[0.5ex]
	$4.5\textless{z}\textless5.5$ & 0.15 & $1.42\pm{0.18}$ & 53 & $1.20\pm{0.07}$ & 22 & 89 \\[0.5ex]
	$5.5\textless{z}\textless7.2$ & 0.01 & $1.02\pm{0.29}$ & 27 & $0.93\pm{0.09}$ & 17 & 41 \\[0.5ex]
	\hline\\[-2.5ex]	
	\end{tabular}
	\end{center}
	%\vglue -5ex
			$\phantom{}$$^{*}$ Fits to the {\msr} assume the form: \\
			$r~(m_{*})/${kpc}$=r_0~{\cdot}~m_{*}^{0.15}$\\
			$\phantom{}$$^{**}$ We correct sizes measured from image stacks due to misalignment of individual galaxy position angles. The corrections are given in the column '\%'.\\
\end{table}
\subsection{Average Size Measurements}
We calculate the average of $r_{1/2,maj}$ in redshift bins to $z\sim7$ by fitting {\ms} distributions assuming:
\begin{equation} r~(m_{*})/\text{kpc}=r_0~{\cdot}~m_{*}^{\alpha} \end{equation}
Where $\alpha$ is the slope, $r_0$ is mass-normalised average of $r$, and $m_{*}\equiv{M}_{*}/${\m} is the ratio of the galaxy's stellar mass to a normalising mass, $m_{0}$.
We choose log({\m})~$=10.1~\text{M}_{\odot}$ because it represents the median mass of the entire ZFOURGE sample at all redshifts.
\begin{figure*}[]
\epsscale{1.2}\plotone{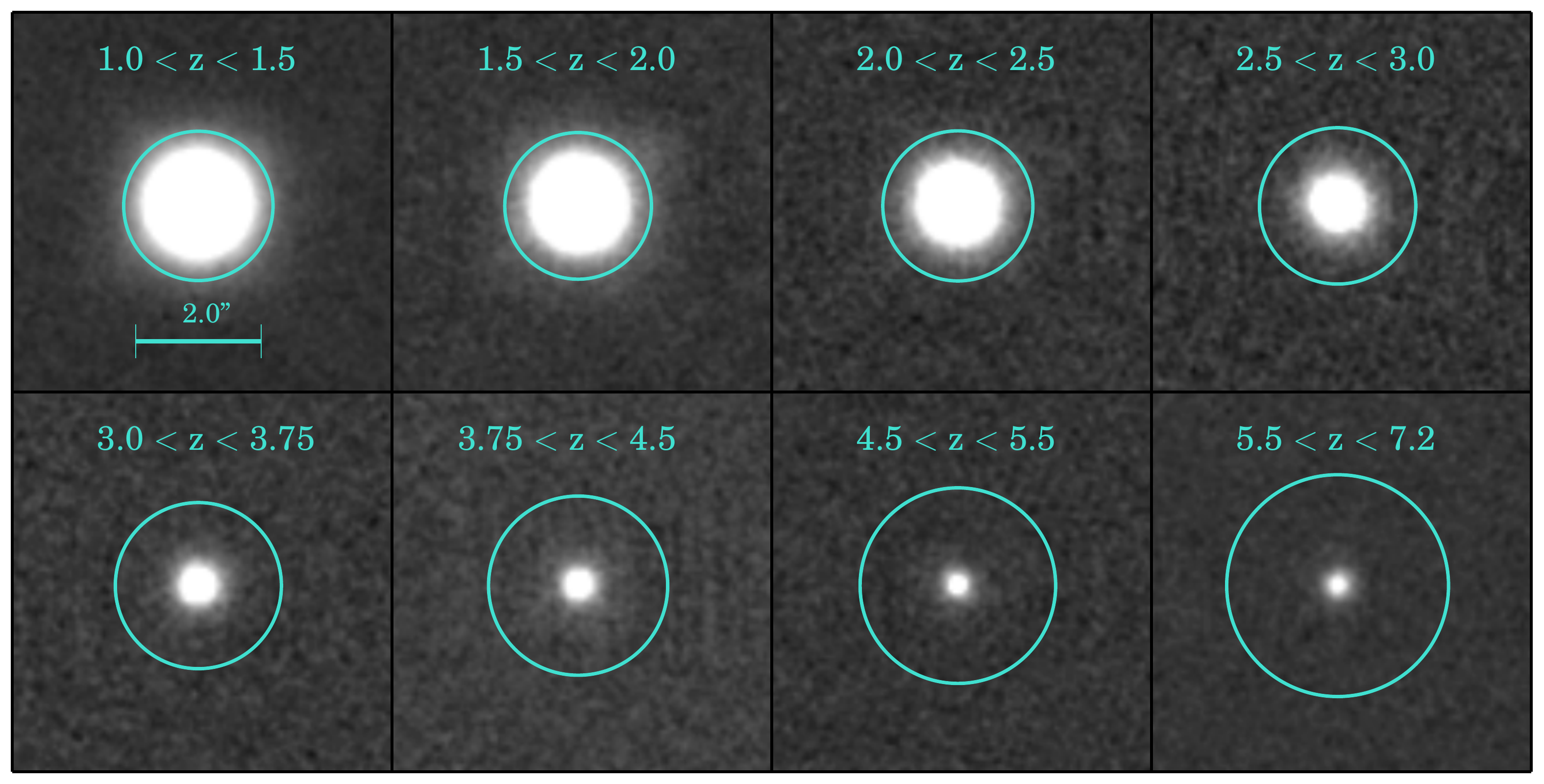}
  \caption{Stacks of galaxy images from {\HST}/F160W CANDELS imaging. Each galaxy image is normalised by its F160W flux before being stacked. The circles in each panel have a diameter of 10{\kpc} at the stacked redshift. Note the apparent decrease in brightness and size of the average galaxy light profiles with redshift.}
  \label{fig:zs}
\end{figure*}

The number of {\sfgs} at fixed mass decreases with increasing redshift; therefore, we fix $\alpha$ and simply fit for $r_0$. 
The value for $\alpha$ is $0.15\pm{0.01}$, which we determine by fitting the {\ms} distributions in the three lowest redshift bins ($1\textless{z}\textless2.5$) and calculating the weighted average of those values.
To estimate the error in $r_0$ we perform 1000 bootstrap calculations in each redshift bin. 
In Figure~\ref{fig:size}, we show the {\ms} distributions to $z\sim7$.

To test the stability of our average sizes we use the full 3D-HST galaxy catalogs, which includes the sample that overlaps with ZFOURGE.
We find that the average sizes do not change; however, the scatter in $r_{1/2,maj}$ increases with increasing redshift for the 3D-HST galaxies.
We do not include their full sample of galaxies in our final analysis because we cannot visually inspect their galaxy SEDs to determine if the photometric redshifts are reliable above $z=4$.
\subsection{Sizes from {\HST}/F160W image stacks}
Individual galaxy sizes may not be reliable because of the low signal-to-noise (S/N) in individual galaxy images, and we test this by measuring average galaxy sizes from image stacks.
We stack galaxy images over the same redshift intervals considered for the {\msr} analysis.
The survey footprints of ZFOURGE and 3D-HST do not perfectly overlap; therefore, the sample size of galaxies for each redshift is larger for the image stack analysis because we use the full ZFOURGE catalogs (see Table~\ref{table:fit}). 

We measure $r_{1/2,maj}$ using the same method as in \citet{2016ApJ...826...60A}, and summarise the steps here.
Individual galaxy stamps are created from CANDELS {\HST}/F160W images.
We shift each stamp so that the central pixel corresponds to the isophotal center of the galaxy determined by SExtractor \citep{1996A&AS..117..393B}.
Masks for each stamp are created using SExtractor, where all objects except the central galaxy are flagged.
Before combining the stamps, each galaxy is normalised by its F160W flux such that the stacks are not biased by bright objects.
Galaxies that are resolved multi-component systems in {\HST} imaging, but were single galaxies in the ZFOURGE catalog are excluded from the stack due to SED uncertainties, such as mass and redshift.

Regardless of the elimination of some galaxies due to resolution issues, each image stack contains more galaxies per redshift bin than the {\ms} distributions (see Table~\ref{table:fit} for the final sample sizes ).
This is advantageous as the S/N of the stacks improves as the square root of the number of images, which should reveal if measuring average sizes at high redshift from the {\ms} distribution is biased by low S/N in individual galaxy images.

In Figure~\ref{fig:zs}, we show the final image stacks for each redshift bin.
In each panel we include an annulus centered on the galaxy that has a radius equal to $10${\kpc} for that redshift (taken as the median redshift of the bin).
It can be seen from the image stacks alone that the light profile of the galaxy decreases with increasing redshift. 

The background subtracted image stacks are fit using GALFIT, assuming a single component {\n} profile.
We estimate errors in the average sizes by bootstrapping the image stacking 1000 times and then refitting them.

We do not rotate individual galaxies so that they are all aligned on their major axis before stacking due to the large uncertainty in position angles (PAs) beyond $z\geq1.5-2$.
However, this may affect size measurements from image stacks.
To understand how measured sizes are affected, we create model galaxies with random sizes and magnitudes chosen from the real distributions of our samples in each redshift bin.
We do this for two sets of models; one which has random PAs and the other fixed PAs (PA=90$\deg$), and then stack them separately.
We find that the sizes measured from image stacks of model galaxies with fixed PAs are $17-22\%$ larger than sizes measured from stacks where the position angle is random at all redshifts.
We apply a correction to the average sizes measured from images stacks equal to the percent size offset in each redshift bin.
The percent corrections, corrected sizes and their errors are shown in Table~\ref{table:fit}.
\begin{figure*}[]
\epsscale{1.18}\plotone{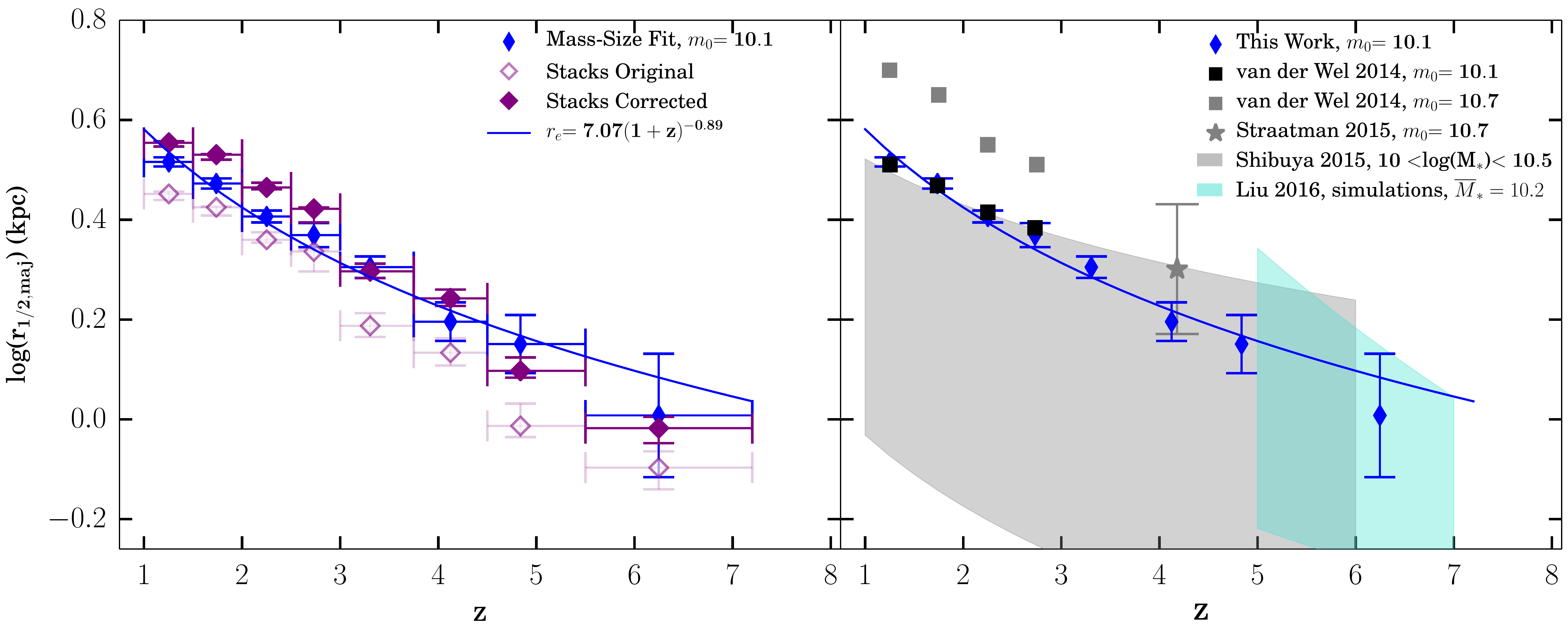}
  \caption{Left Panel: The evolution of star-forming galaxy sizes to $z~\sim7$. We show average sizes  via fitting the {\msr}, with log({\m})~$=10.1${\s}, as blue diamonds, and position angle corrected (uncorrected) average sizes from image stacks as solid (open) purple diamonds. Averages calculated via fitting the {\msr} are consistent with corrected average sizes measured from image stacks. We fit the size evolution (blue line) and find $r_e=7.07(1+z)^{-0.89\pm{0.01}}${\kpc} for {\sfgs} with {\M}$\textgreater10$. Right Panel: We compare our fit to the size evolution of {\sfgs} with previous mass complete studies. Grey points are from \citet{2014ApJ...788...28V} who use log({\m})~$=10.7${\s} when fitting the {\msr}. When we fit their data below $z=3$ using log({\m})~$=10.1${\s}, we measure consistent sizes shown as black points. The grey star from \citet{2015ApJ...808L..29S} is the median circularised size. The grey contour represents the median circularised sizes and 16th and 84th percentile distribution of individual sizes for {\sfgs} with $10\textless${\M}$\textless10.5$ \citep{2015ApJS..219...15S}. The turquoise contour represents the median and percentile distribution of simulated disk sizes from \citet{2016arXiv160800819L}.}
  \label{fig:del}
\end{figure*}
\begin{figure*}[]
\epsscale{1.1}\plotone{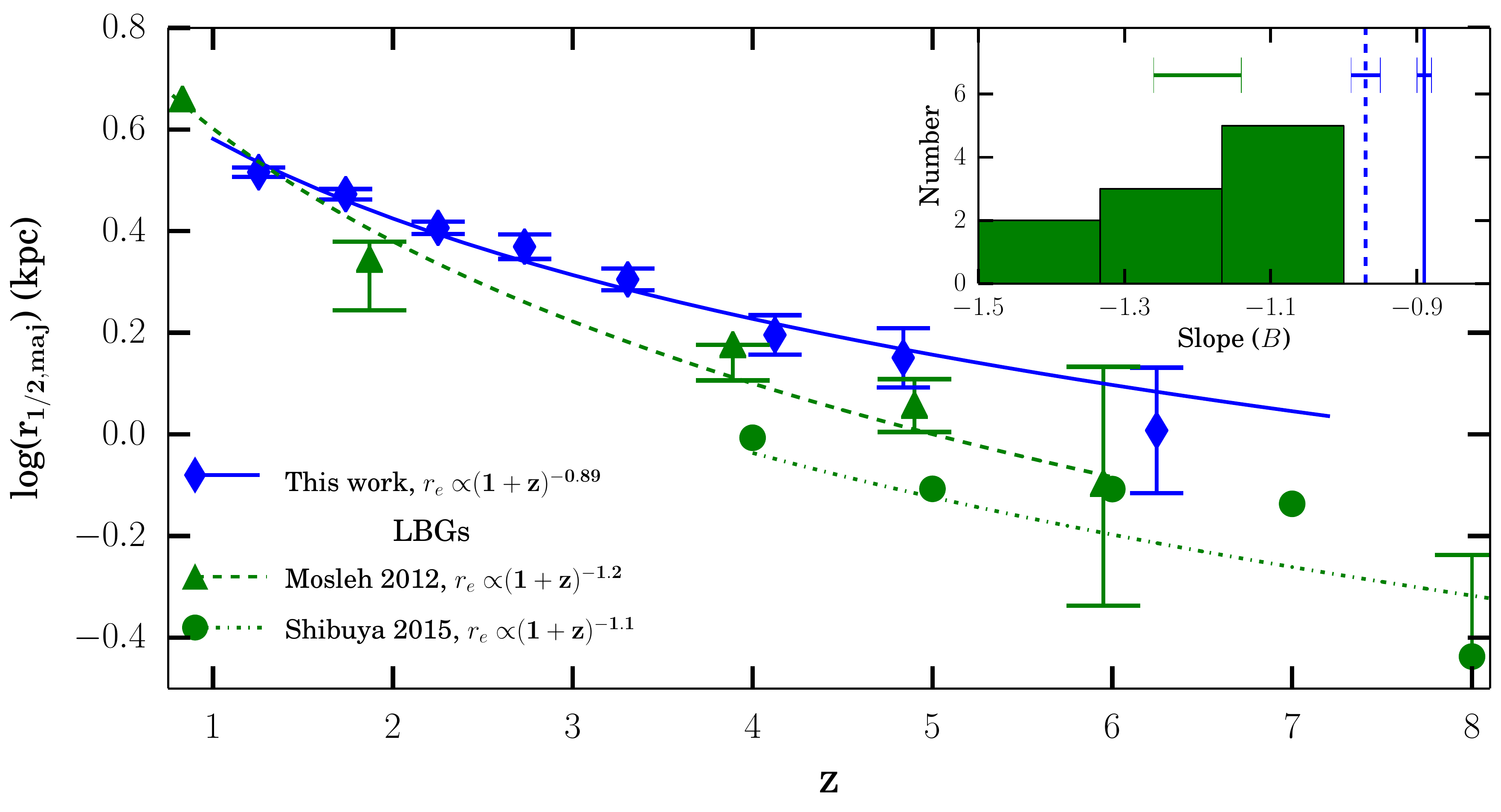}
  \caption{The evolution of star-forming galaxy sizes from our mass complete sample compared to the size evolution derived for LBGs. The green triangles and dashed line are the median circularised sizes and fit to the size evolution of LBGs with $9.5\textless${\M}$\textless10.4$ \citep{2012ApJ...756L..12M}. The green dot-dashed line is our fit to the best-fit sizes (green circles) from the luminosity$-$size relation of LBGs with $L=1L_{*, z=3}$ from \citep{2015ApJS..219...15S}. Both green lines are offset in size normalisation due to mass selection effects; however, it is important to note that the LBG size evolution slopes are different, which indicate a different growth rate from typical {\sfgs}. This is further shown in the inset histogram where we include measured slopes for LBGs from previous works listed in \citet[][Table 7]{2015ApJS..219...15S}. The average of those slopes is $1.20\pm{0.06}$ (green error bars), which is significantly steeper compared to our slopes determined from the {\ms} relation (blue line with errors) and the median of circularised sizes (blue dashed line with errors).}
  \label{fig:lbg}
\end{figure*}
\section{Results}
\label{sec:res}
For the first time, we use a mass-complete ({\M}$\textgreater10$) sample of {\sfgs} to extend the {\msr} to $z\sim7$.
We show in Figure~\ref{fig:del} (left panel), the evolution of average galaxy sizes at fixed mass measured via fitting the {\msr} and from image stacks.
At $z\sim7$, the sizes of star-formers are extremely compact with $r_{1/2,maj}\sim1${\kpc}.

In Figure~\ref{fig:del} (left), we compare the average sizes measured via fitting the {\msr} (filled blue diamonds) to those measured from non-corrected image stacks (open purple diamonds), and we find that the stacked sizes are much smaller.
After applying a correction to compensate for the varying PAs of individual galaxies, the sizes become consistent with those measured via fitting (See Figure~\ref{fig:del}, filled blue and purple diamonds).
Therefore, we conclude that fitting the {\msr} of individual galaxies is not affected by low S/N galaxies, and both methods produce reliable sizes.

We test whether we are measuring a true size evolution or simply measuring the peak of $z=5-7$ galaxy light profiles by red-shifting the individual sizes and F160W magnitudes of our $z\sim1.25$ galaxy sample to resemble those of $z\sim7$ galaxies.
The model galaxies have random axis ratios, position angles, and are flux weighted before combined.
In addition, we include Poisson noise to the image stacks before they are fit with GALFIT.
When we use the $z\sim7$ angular to kpc conversion, we are able to recover the original $z\sim1.25$ galaxy sizes, confirming the robustness of our average size measurement at $z\sim7$.

We fit a power-law relation to the average sizes measured via fitting the {\msr} from {\zr}, using $r_e=A(1+z)^B${\kpc}. 
We find $r_e=7.07(1+z)^{-0.89\pm{0.01}}${\kpc} for {\sfgs} with {\M}$\textgreater10$.
In both panels of Figure~\ref{fig:del}, we show our fit as a blue line. 
\section{Discussion}
\label{sec:dis}
\subsection{Mass Complete Studies}
At $1\textless{z}\textless3$, \citet{2014ApJ...788...28V} measured the size evolution of star-formers with {\M}$\textgreater9.5$ by fitting the {\msr} using log({\m})~$=10.7${\s}.
When we fit their size-mass data using our mass-limit and value of log({\m})~$=10.1${\s}, we recover our size results.
This can be seen in Figure~\ref{fig:del} (right panel), where we show their original results (grey points) and the refitted normalised sizes (black points).
However, \citet{2014ApJ...788...28V} found a more shallow size evolution with $B=-0.75$, compared to what we find, $B=-0.89\pm{0.01}$.
Because our measured sizes are consistent, we attribute the discrepancy in measured size evolution to the redshift limit of their study (see Figure~\ref{fig:del}, right panel).

At $z\sim4$, \citet{2015ApJ...808L..29S} found that the median circularised size of {\sfgs} with {\M}$\textgreater10.6$ is $r_{e}=2\pm0.60${\kpc}.
While the mass limit of their sample is higher, and they use circularised sizes, our $z\sim4$ size is consistent within the errors.
However, their measured size evolution, determined from their $z\sim4$ median size and the median sizes of lower redshift galaxies from \citet{2014ApJ...788...28V}, follows $r_e\propto(1+z)^{-0.72}${\kpc}, and is also more shallow than what we find.
Because of the large errors in their median size, due the scatter in individual sizes, they are not able to reliably constrain the redshift dependency of star-forming galaxy sizes.

\citet{2015ApJS..219...15S} calculated the median circularised sizes of {\sfgs} at $0\textless{z}\textless6$ with $10\textless${\M}$\textless10.5$ (grey contour in Figure~\ref{fig:del}, right), however they use a Salpeter IMF when fitting galaxy SEDs to estimate stellar masses.
Regardless of the difference in methodology, our average sizes are consistent within the 16th and 84th percentile distribution of their median sizes.
The large scatter they measured is due to the fact that they are taking the median over $0.5$~dex in mass, and because of the {\msr} more massive galaxies will be larger.
The size evolution traced by their data is also consistent with our result at $z\textgreater2$.

We find that our results are consistent with recent cosmological galaxy disk simulations from the Dark-ages Reionization And Galaxy-formation Observables from Numerical Simulation (DRAGONS) series \citep{2016arXiv160800819L}, which indicates that high redshift star-formers are indeed disk dominated. 

\subsection{LBG Studies}
By far the largest population of star-formers to be studied at $z\textgreater4$ are LBGs.
While the size evolution of these galaxies is well documented from $4\textless{z}\textless12$~\citep[e.g.][]{2012ApJ...756L..12M,2013ApJ...777..155O,2015ApJ...808....6H,2015ApJS..219...15S}, samples of galaxies are selected via filter dropout techniques, and are not mass-complete.
Theoretically, the typical luminosity range considered (i.e., L$=0.3-1L_{*,z=3}$) spans a range of masses ($9.2\textless${\M}$\textless10.7$ at $5\textless{z}\textless6$, from \citet{2015ApJS..219...15S}, Figure~1).
However, as seen in \citet[][]{2012ApJ...756L..12M}, the median mass at any redshift is less than {\M}$=10$.
This is supported by the higher number densities found for LBGs at $z\sim7$, N=0.8 arcmin$^{-2}$ \citep{2012ApJ...754...83B}, compared to our mass complete sample, N=0.1 arcmin$^{-2}$.
Therefore, LBGs may not represent the entire population of star-formers at $z\textgreater6$.

To understand if LBGs represent a special subset of highly {\sfgs}, we attempt to compare our size results with current LBG literature.
However, LBG studies employ different methodologies than used in our work concerning IMF choice, size definition, and quantification of typical galaxy size.
Therefore, we have adjusted our data where appropriate to be able to more directly compare our results.

When we compare the median circularised sizes of LBGs with $9.5\textless${\M}$\textless10.4$ at $z\sim4, 5,$ and $6$ measured by \citet{2012ApJ...756L..12M} to our average sizes, we find that they are consistent within $2\sigma$, given their large error bars (Figure~\ref{fig:lbg}, green diamonds).
However, their fit to the size evolution of these galaxies, $r_e=A(1+z)^B${\kpc}, has a steeper slope ($B=-1.20$) compared to our mass-complete sample ($B=-0.89$), shown in Figure~\ref{fig:lbg}, green dashed line.
While we expect LBGs to have smaller sizes due to the mass range of their samples, the steeper slope indicates a different size growth rate.
To understand if this discrepancy is driven by a fundamental difference in galaxy samples, or simply due to a difference in methodology, we adjust our data by converting stellar masses from Chabrier to Salpeter, and then calculate the median circularised sizes.
The size evolution we measure becomes more steep ($B=-0.97\pm{0.02}$); however, it is inconsistent with $B=-1.20$.

The steep value found for $B$ in \citet{2012ApJ...756L..12M} is supported by other studies that examine the size evolution of LBGs using median sizes, as well as sizes from the luminosity$-$size relation \citep[e.g.][]{2004ApJ...616L..79B,2013ApJ...777..155O,2010ApJ...709L..21O,2015ApJ...808....6H,2015ApJS..219...15S}.
For example, \citet{2015ApJS..219...15S} measured the size evolution of LBGs in luminosity bins of $L=0.12-10L_{*, z=3}$ using averages, medians, and modes of circularised sizes, as well as, fits to the luminosity$-$size relation.   
Galaxies with $L=1L_{*, z=3}$ span different mass ranges depending on their redshift making it difficult to directly compare average sizes measured from the luminosity$-$size relation to those from the {\msr}. 
The size evolution \citet{2015ApJS..219...15S} measured, based on fitted sizes from the luminosity$-$size relation, is more steep with $r_e\propto(1+z)^{-1.2}${\kpc}, compared what we find using the {\msr}.
When we compare slopes based on median circularised sizes, we find B=$-0.97\pm{0.02}$, while they find that $B=-1.15\pm{0.07}$ for LBGs.
Furthermore, if we average the values of $B$ from other studies listed in \citet[][Table 7]{2015ApJS..219...15S} (parametric fits only, but determined from a range of statistics), we find $\bar{B}=-1.20\pm{0.06}$, which is steeper (by $3.6\sigma$) than our slope determined using median sizes (see inset of Figure~\ref{fig:lbg}).

By selecting LBGs within our sample and comparing their {\msr} to non-LBG galaxies, we could determine if LBGs represent a special compact population of star-formers. 
However, ZFOURGE is a K-band selected sample, and we are not sensitive to or biased towards selecting low mass LBGs. 
If we attempt to select LBG-like galaxies from our sample by selecting dust-free galaxies, using a color cut of U$-$V and V$-$J $\textless1$, or by further extending our mass limit below our completion threshold of {\M}$=10$, we find a size evolution consistent with typical galaxies and inconsistent with LBGs.

While we expect an offset in the size evolution of LBGs, due to their lower median masses compared to our sample, the different growth rates we find is interesting.
To date, samples of LBGs are not mass complete; therefore, these galaxies likely represent a special subsample of highly star-forming and compact galaxies.
This is not surprising given diversity in dust content, UV magnitudes, and star-formation that \citet{2014ApJ...787L..36S} showed for star-formers with {\M}$\textgreater10.6$ at $z=3-4$.
\section{conclusions}
\label{sec:con}
For the first time, we trace the size evolution of a complete sample of {\sfgs} to $z\sim7$, and confirm that the {\msr} exists at this redshift.
The sizes of {\sfgs} at $z\sim7$ are extremely compact and have $r_{1/2,maj}\sim1${\kpc}.
We find a steeper size evolution of the form $r_e=7.07(1+z)^{-0.89\pm{0.01}}${\kpc} compared to previous mass compete works, but different to the size evolution found for LBGs.
\\
%\acknowledgments

\ack

%\bibliographystyle{apj}
%\bibliography{mybibliography}

\end{document}